\newcommand*{\ATLASLATEXPATH}{./}
\documentclass[cernpreprint,txfonts,UKenglish,texlive=2011,PAPER,atlasstyle=true]{\ATLASLATEXPATH atlasdoc}
 \pdfoutput=1

\usepackage[csquotes=false,biblatex=true,minimal]{\ATLASLATEXPATH atlaspackage}

\usepackage{\ATLASLATEXPATH atlascontribute}

\usepackage{\ATLASLATEXPATH atlasphysics}
\usepackage{\ATLASLATEXPATH atlasphysics}
\usepackage{\ATLASLATEXPATH atlascontribute}
\usepackage{\ATLASLATEXPATH atlasbiblatex}
\usepackage{xspace}
\usepackage{tikz}
\usepackage{siunitx}
\usepackage{multirow}
\usepackage{caption}
\usepackage{subcaption}
\usepackage[figuresright]{rotating} 
\usepackage{lscape}
\usepackage{longtable}

\graphicspath{{logos/}{figures/}}

\usepackage{WW1JetPaper-defs}
\usepackage{WW1JetPaper-results}

\hypersetup{pdftitle={ATLAS draft},pdfauthor={The ATLAS Collaboration}}

\AtlasTitle{Measurement of $W^+W^-$ production in association with one jet in proton--proton collisions at $\mathbf{\sqrt{s} =8}$~\TeV\ with the ATLAS detector}

\author{The ATLAS Collaboration}

\date{\today}

\AtlasRefCode{STDM-2015-24}

\PreprintIdNumber{CERN-EP-2016-186}

 \AtlasJournal{Phys.\ Lett.\ B.}
 \AtlasJournalRef{\PLB 763 (2016) 114}
 \AtlasDOI{10.1016/j.physletb.2016.10.014}

\AtlasAbstract{%
 The production of $W$ boson pairs in association with one jet in $pp$
collisions at $\sqrt s =\ 8$~\TeV\ is studied using data corresponding to an integrated luminosity of 20.3 fb$^{-1}$ collected by the ATLAS detector during 2012 at the CERN Large Hadron Collider. The cross section is measured in a fiducial phase-space region defined by the presence of exactly one electron and one muon, missing transverse momentum  and exactly one jet with a transverse momentum above 25~\GeV\ and a pseudorapidity of $|\eta|<4.5$. The leptons are required to have opposite electric charge and to pass transverse momentum and pseudorapidity requirements. The fiducial cross section is found to be $\sigma^{\text{fid,1-jet}}_{WW}=136\pm6\left(\text{stat}\right)\pm14\left(\text{syst}\right)\pm3\left(\text{lumi}\right)$~fb. In combination with a previous measurement restricted to leptonic final states with no associated jets, the fiducial cross section of $WW$ production with zero or one jet is measured to be $\sigma^{\text{fid,$\leq$1-jet}}_{WW}=511\pm9\left(\text{stat}\right)\pm26\left(\text{syst}\right)\pm10\left(\text{lumi}\right)$~fb. The ratio of  fiducial cross sections in final states with  one and zero jets is determined to be $0.36\pm0.05$. Finally, a total cross section extrapolated from the fiducial measurement of $WW$ production with zero or one associated jet is reported.  The measurements are compared to theoretical predictions and found in good agreement.
}

 \AtlasCoverSupportingNote{\centering Support note on CDS}{https://cds.cern.ch/record/2114862/}
 \AtlasCoverAnalysisTeam{Valerio Dao, Evangelos Kourlitis, Yichen Li, Kristin Lohwasser, Philip Sommer, Christian Weiser, Yusheng Wu, Li Yuan, Zhiqing Zhang}

 \AtlasCoverEdBoardMember{John Hobbs, Chariclia Petridou~(chair), Emmanuel Sauvan}

 \AtlasCoverEgroupEditors{atlas-STDM-2015-24-editors@cern.ch}

 \AtlasCoverEgroupEdBoard{atlas-STDM-2015-24-editorial-board@cern.ch}

\begin{document}

\maketitle




\section{Introduction}
\label{sec:introduction}

The measurement of the production of two $W$ bosons is a crucial test of the non-Abelian gauge structure of the electroweak theory of the Standard Model (SM). The increasing precision of the experimental measurements at the LHC has elicited improved  theoretical descriptions of the process. Progress has been made to extend the next-to-leading-order (NLO)~\cite{Campbell:2011bn} calculation of $pp\rightarrow W^+W^-$ production to include next-to-next-to-leading-order (NNLO) effects~\cite{Gehrmann:2014fva} in perturbative quantum chromodynamics (QCD). A separate calculation of the loop-induced, non-resonant $gg\rightarrow W^+W^-$ production process has been made available at order $\mathcal{O}(\alphas^3)$~\cite{Caola:2015rqy} in the strong coupling constant $\alphas$. Resonant $WW^*$ production via the exchange of a Higgs boson has been calculated to order $\mathcal{O}(\alphas^3)$~\cite{HiggsXS} and $\mathcal{O}(\alphas^4)$~\cite{Anastasiou:2015ema}. These predictions can be summed to give an updated prediction for the total cross section of $65.0^{+1.2}_{-1.1}$~pb as further detailed in Section~\ref{sec:results}. In addition to these new calculations, fully differential NNLO predictions~\cite{Grazzini:2016ctr} have become available, as have dedicated NLO predictions for jet-associated $WW$ production~\cite{Dittmaier:2009un,Campbell:2015hya} with up to three jets~\cite{Cordero:2015hem}. The resummation of logarithms arising from a selection on the number of jets has been presented at next-to-next-to-leading-logarithm (NNLL) accuracy in Refs.~\cite{Meade:2014fca,Grazzini:2015wpa}. It is therefore interesting to study $WW$ production in association with jets to confront these calculations with experimental data from the LHC. 

A measurement of the jet multiplicity in $WW$ events at the CDF experiment was published in Ref.~\cite{Aaltonen:2015hza}. At the LHC, the CMS Collaboration 
has included $WW$ production in association with one jet in their measurement of the total $WW$ production cross section at $\sqrt{s}=8$~\TeV~\cite{Khachatryan:2015sga}, but has not published dedicated fiducial cross sections of jet-associated $WW$ production. 

This letter presents a measurement of the fiducial cross section of $WW$ production using the decay chain $W^+W^-\rightarrow e^{\pm}\nu_e\mu^{\mp}\nu_\mu$ in final states with one associated hadronic jet, further referred to as 1-jet final state. The fiducial region is defined using stable particles at the generator level and is chosen to match the experimental selection as closely as possible.

Only events with exactly one reconstructed jet are selected for the analysis, while events with a larger number of jets suffer from a large background from top-quark production and are not considered.
The selected $WW$ candidate event sample is corrected for background processes, detection efficiencies and resolution effects, and the cross section of $WW$+1-jet production is extracted for the fiducial phase-space region. 
The results are combined with a previous measurement reported in Ref.~\cite{Collaboration:2010127}  restricted to final states without any reconstructed jets,  referred to as 0-jet final state. The  fiducial $WW$+$\leq$1-jet cross section and the ratio $R_1$ of the fiducial $WW$+1-jet and fiducial $WW$+0-jet cross sections are determined and compared to different theoretical predictions. 
The measurement therefore extends the fiducial phase space of the previous measurement  of the $WW$ production cross section.


\section{Data and Monte Carlo samples}
\label{sec:samples}
\label{sec:MCSamples}

The ATLAS detector~\cite{detectorpaper} is a general-purpose detector measuring collisions at the Large Hadron Collider~(LHC) with coverage over the full azimuthal angle $\phi$. It consists of an inner detector surrounded by a 2 T solenoid to measure tracks with pseudorapidities of $|\eta|<2.5$\footnote{ATLAS uses a right-handed coordinate system with its origin at the nominal interaction point (IP) in the centre of the detector and the $z$-axis along the beam pipe. The $x$-axis points from the IP to the centre of the LHC ring, and the $y$-axis points upward. Cylindrical coordinates $(r,\phi)$ are used in the transverse plane, $\phi$ being the azimuthal angle around the $z$-axis. The pseudorapidity is defined in terms of the polar angle $\theta$ as $\eta=-\ln\tan(\theta/2)$. The transverse energy is computed as $\et = E \cdot \sin\theta$, while the radial distance between two objects is defined as $\Delta R = \sqrt{\left(\Delta\eta\right)^2+\left(\Delta\phi\right)^2}$.
}, electromagnetic and hadronic calorimeters to provide energy measurements for  $|\eta|<4.9$, and a muon spectrometer with a toroidal magnetic field to detect muons with $|\eta|=2.7$. 
A three-level trigger system selects events to be read out.

The measurement uses data collected with the ATLAS experiment during the 2012 data-taking period. Only runs with stable proton beams colliding at $\sqrt{s}=8\,\TeV$ are used in which all relevant detector components were functional. This data sample corresponds to an integrated luminosity of 20.3\,\ifb\ determined with an uncertainty of $\pm1.9\%$ and derived from beam-separation scans performed in November 2012~\cite{Aaboud:2016hhf}.

The analysis relies on event simulation to correct the measured event yields for experimental effects and for the study of background processes. Different simulated event samples are used to model the signal
from the individual production mechanisms: $q\bar{q}\rightarrow W^+W^-$ events are simulated using the \Powheg~1.0 generator~\cite{Nason:2004rx, Frixione:2007vw, Alioli:2010xd, Alioli:2008gx,Melia:2011tj}, which is interfaced to \Pythia~8.170~\cite{newpythia8}; for the non-resonant $gg$-induced $WW$ signal the \ggtwoww\ program (version 3.1.3)~\cite{gg2ww}  is employed and interfaced to \herwig~6.5/\Jimmy~4.31~\cite{Herwig,Butterworth:1996zw};  resonant $WW^*$ production via a Higgs boson with a mass of $m_H =125$~\GeV\ is modelled using \Powheg+\Pythia~8.170. The three event samples are simulated using the  CT10 NLO~\cite{cteq10} parton distribution function (PDF). 
Photon radiation is modelled using \Photos~\cite{Golonka:2005pn}.
The parameter tune used for the underlying event is AU2~\cite{ATLAS:2012uec}.
 The event samples are normalised to a cross section times branching ratio of 5.58~pb ($q\bar{q} \rightarrow W^+W^-$~\cite{Campbell:2011bn}), 0.153~pb (non-resonant $gg \rightarrow W^+W^-$~\cite{gg2ww}) and 0.435~pb ($gg \rightarrow  H \rightarrow W^+W^-$~\cite{HiggsXS}). 
The sum of these contributions corresponds to a total $WW$ cross-section of $58.7^{+4.2}_{-3.8}$~pb where the uncertainties are due to scale and PDF uncertainties in the  cross section calculations. 
For additional studies a sample of simulated $q\bar{q} \rightarrow W^+W^-$ events produced with \textsc{MC@NLO}~\cite{Frixione:2007vw} and \textsc{Jimmy}~\cite{Herwig,Butterworth:1996zw} using the AUET2 tune~\cite{ATLAS:2010osr} and the CT10 PDF is used.

Production of pairs of top quarks, $s$-channel single top-quark production and $W$-associated top-quark production are modelled with the \Powheg+\Pythia~6  generator with the AU2~\cite{ATLAS:2012uec} tune. Single top-quark production in the $t$-channel is described by the \textsc{Acer} 3.7~\cite{Kersevan:2004yg} MC generator interfaced to \Pythia~6~\cite{pythia} with the AUET2B tune~\cite{ATL-PHYS-PUB-2011-009}. These events samples are normalised to the respective NNLO+NNLL calculations~\cite{Czakon:2011xx,Kidonakis:2010ux,Kidonakis:2011wy,Kidonakis:2010tc} to obtain the relative contribution to the total top-quark background, whose overall normalisation is determined from data as detailed in Section~\ref{sec:backgrounds}. 

Background from $W$ and $Z$ boson production is modelled using \Alpgen\ 2.14~\cite{Mangano:2002ea} interfaced to \Pythia~6 and normalized to NNLO calculations~\cite{Catani:2007vq} where needed. The AUET2 tune is used for the underlying event. The diboson background processes $WZ$ and $ZZ$ are generated using the same settings as employed for the simulated $q\bar{q} \rightarrow W^+W^-$ event samples. The production of a $W$ boson and a virtual photon ($\gamma^*$) is generated using the \textsc{Sherpa} generator (version 1.4.2)~\cite{Sherpa}. For $W\gamma$ production \Alpgen+\Herwig+\Jimmy\ is employed.

In all simulated event samples, additional  $pp$ collisions accompanying the hard-scatter interactions (pile-up) are modelled  by overlaying minimum-bias events generated using \Pythia~8. To simulate the detector response, the generated events are passed through a detailed simulation of the ATLAS detector~\cite{Aad:2010ah} based on \textsc{Geant4}~\cite{Agostinelli:2002hh} or \textsc{Geant4} combined with a parameterised calorimeter simulation~\cite{ATLAS:1300517}. 

\section{Object reconstruction and event selection}
\label{sec:selection}

Events are selected using reconstructed jets, electrons, muons and missing transverse momentum. The selection follows closely the one in  Ref.~\cite{Collaboration:2010127} to facilitate the combination with the $WW$+0-jet final state. Electrons and muons are identified based on tracks in the inner detector matched either to energy deposits in the electromagnetic calorimeter or combined with tracks in the muon spectrometer, respectively. Electrons are reconstructed within $\left|\eta\right|<2.47$ excluding the transition region between barrel and endcap calorimeters of $1.37<\left|\eta\right|<1.52$. Muons are required to lie within $\left|\eta\right|<2.4$. The same reconstruction and identification requirements as in Ref.~\cite{Collaboration:2010127} are used, resulting in an event sample with minimal contributions from backgrounds due to particles misidentified as leptons, particularly from $W+$jets, \mj\ and $W\gamma$ events. For the selection of $WW$ candidate events, the presence of exactly two isolated, oppositely charged leptons $(\ell,\ell')$ with transverse momenta of $\pt^{\ell} > 25$~\GeV\ and $\pt^{\ell'} > 20$~\GeV\ is required. Only final states with one electron and one muon are used. Events with additional leptons with $\pt>$ 7~\GeV\ are rejected, which helps to suppress other diboson processes with more than two leptons. It is required that at least one of the leptons has met an online single-lepton selection or both  have passed a dilepton trigger with reduced thresholds and less stringent object identification criteria. This setup has an efficiency of 99\%-100\% with respect to the offline lepton selection.

Jets are formed using calibrated topological clusters of energy~\cite{bib:topo} reconstructed in the calorimeters using the anti-$k_t$ algorithm~\cite{bib:antikt2} with radius parameter $R=0.4$. Further corrections to the jet energy are applied based on  simulation~\cite{JES} and are followed by a pile-up suppression~\cite{cacciari:2007fd}. 
Jets are required to have $\pt>25$~\GeV\ and $\left|\eta\right| < 4.5$. 
More than 50\% of the scalar sum of the \pt\ of all tracks contained within $\Delta R=0.4$ of the jet axis is required to be from tracks associated with the primary vertex to suppress contributions from additional $pp$ interactions in the event~\cite{atlas-conf-2013-083} if the jet satisfies $\pt<50$~\GeV\ and $\left|\eta\right|<2.4$.
Only events with exactly one jet meeting the above criteria are selected. Jets containing $b$-hadrons (so-called $b$-jets) are identified within the central region of the detector, $\left|\eta\right|<2.5$, using a multivariate approach~\cite{atlas-conf-2011-102,atlas-conf-2014-046} with an efficiency of 85\%. To reduce the background from top-quark production, events containing $b$-jets with $\pt>20$~\GeV\ and within $\left|\eta\right| < 2.5$ are rejected.

Selection requirements on the missing transverse momentum in the candidate events are used to reduce the contribution of events from 
$Z/\gamma^*\rightarrow\tau\tau$ (Drell--Yan) production where both $\tau$-leptons decay leptonically.
Missing transverse momentum is reconstructed from the vector sum of the transverse momenta of identified particles~\cite{Aad:2012re} to which either reconstructed jets and calorimetric depositions not associated with any particle are added. Missing transverse momentum induced by mismeasurements of the energy of leptons is further reduced in the calorimeter-based measurement by projecting the missing transverse momentum \met~onto nearby leptons, to calculate the so-called relative missing transverse momentum \metrel. A lepton is considered nearby if the azimuthal separation to the $\vec{E}_{\mathrm{T}}^{\mathrm{miss}}$~direction is small, $\Delta\phi(\met,\ell)<\pi/2$, and only in this case, \met\ is modified to yield $\metrel=\met \times \sin(\Delta\phi(\met,\ell))$, otherwise $\metrel=\met$. The relative missing transverse momentum is required to be $\metrel>15$~\GeV. An additional track-based measure of the missing transverse momentum (\ptmiss) is constructed by adding the momenta of tracks associated with the primary vertex to the vector sum of the transverse momenta of identified electrons and muons. By construction,  \ptmiss\ is less sensitive to energy deposits from additional interactions and it is required to be $\ptmiss>20$~\GeV. To further reduce the sensitivity to fluctuations in either of the missing transverse momentum variables used, the azimuthal separation between  $\vec{E}_{\mathrm{T}}^{\mathrm{miss}}$\ and  $\vec{p}_{\mathrm{T}}^{\mathrm{miss}}$ must satisfy $\Delta\phi(\met,\ptmiss) < 2.0$. 

The invariant mass of the two selected leptons, $m_{\ell\ell}$, is required to be greater than 10~\GeV\ to suppress contributions from misidentified leptons produced in multijet and $W$+jets events. Apart from the requirements on the jets and $\Delta\phi(\met,\ptmiss)$, this event selection is identical to the one employed in Ref.~\cite{Collaboration:2010127}.

\section{Determination of backgrounds}
\label{sec:backgrounds}

The experimental signature of exactly one electron and one  muon with opposite electric charge, and  missing transverse momentum can be produced by a variety of SM processes which are treated as backgrounds. Top quarks decay almost exclusively to a $b$-quark and a $W$ boson. This makes $t\bar{t}$ and single top-quark production the dominant background to $WW$ production, in particular for events with jets in the final state. 
The background yield from top-quark production is determined using a method proposed in Ref.~\cite{ATLAS:2014aga}. The event yield is extrapolated from a control sample enriched in events from top-quark production.
It is defined by the nominal selection requirements but must contain exactly one identified $b$-jet with $\pt>25$~\GeV\ and within $|\eta|< 2.5$, instead of requiring the absence of identified $b$-jets. 
The distribution of the transverse momentum of the $b$-jet in the control sample is shown in Figure~\ref{fig:invmass-jetpta}(a). The data is used to constrain the large experimental and theoretical uncertainties shown by the error bands. 
The factor to extrapolate from this control sample to the signal sample is determined as the ratio of jets passing or failing the $b$-jet requirement in additional control samples, defined by the presence of two jets, at least one of which passes passes the $b$-tag requirement.
Systematic effects resulting from the choice of the control sample are corrected for by an additional factor estimated from simulated event samples.
The correction introduces  experimental systematic uncertainties of $\pm3.1\%$, mainly from the uncertainty in the jet energy scale. Theoretical uncertainties are found to amount to $\pm2.5\%$ and are dominated by differences in simulated $t\overline{t}$ event samples produced with \textsc{Powheg} and \textsc{MC@NLO}, and uncertainties in the $Wt$ production cross section. Statistical uncertainties from the limited size of the control samples in data and simulation introduce an uncertainty of $\pm3.5\%$, resulting in an overall precision in the estimated top-quark background yield of $\pm5.2\%$.

The estimation of the remaining background processes closely follows the methodology described in Ref.~\cite{Collaboration:2010127}.
Data-driven estimates of the yields of $W$+jets and multijet production are determined in an event sample in data that is selected with relaxed identification and isolation criteria for the leptons. The composition of this event sample with genuine and misidentified leptons can be inferred using the probabilities of genuine and misidentified leptons selected with the relaxed criteria to satisfy the nominal lepton selection criteria.
The yield of background from Drell--Yan production is obtained from a simultaneous fit of the distribution of simulated event samples to the $\Delta\phi(\met,\ptmiss)$ distribution of the data in the signal region and in a control sample, defined by a selection of $5~\GeV<\ptmiss<20$~\GeV\ and no selection on $\Delta\phi(\met,\ptmiss)$. The yields of the diboson processes, $WZ$, $ZZ$ and  $W\gamma$ production, are determined using  simulation and are normalised to NLO predictions~\cite{Campbell:2011bn}. The uncertainties assigned to the NLO predictions are inflated to cover differences from the calculations in Refs.~\cite{Grazzini:2015hta,Grazzini:2016swo}. For $W\gamma$ production a $K$-factor is calculated from Ref.~\cite{Grazzini:2015nwa} and applied to the NLO prediction. 

The observed data and the estimated signal and background yields are summarised in Table~\ref{tab:Yields}. Half of the events selected in data are estimated to originate from background processes, where top-quark production represents the largest contribution. The transverse momentum distribution of the selected jet after the final event selection is shown in Figure~\ref{fig:invmass-jetptb}(b), where data is shown together with the simulated $WW$ signal events and the estimated background yields. Good agreement between the data and the estimated yields is observed for the selected $WW$+1-jet candidate sample.

\begin{table}
\centering
  \centering
    \begin{tabular}{ll@{ ~$\pm$~ }l@{ ~$\pm$~ }ll@{ ~$\pm$~ }l@{ ~$\pm$~ }l}
      \hline \hline
      Process & \multicolumn{3}{c}{$WW$+1-jet} & \multicolumn{3}{c}{$WW$+0-jet} \\
      \hline
      Observed Events &  \multicolumn{3}{l}{3458} & \multicolumn{3}{l}{5067} \\ 
      \hline
      Total expected events 
              & 3310 & 50 & 340 & 4420 & 30 & 320 \\
      (Signal + background) \\
      \hline
      $WW$ signal
              & 1490 & 10 & 330 & 3240 & 10 & 280 \\
      \hline
      Top Quark
              & 1236 & 43 & \hphantom{0}49 & \hphantom{0}609 & 18 & \hphantom{0}52 \\
      $W$+jets
              & \hphantom{0}121 & 15 & \hphantom{0}50 & \hphantom{0}250 & 20 & 140 \\
      Drell--Yan
              & \hphantom{0}267 & 12 & \hphantom{0}49 & \hphantom{0}175 & \hphantom{0}3 & \hphantom{0}18 \\
      Other diboson
              & \hphantom{0}195 & \hphantom{0}5 & \hphantom{0}53 & \hphantom{0}150 & \hphantom{0}4 & \hphantom{0}30 \\
      \hline
      Total background
      & 1820 & 50 & 100 & 1180 & 30 & 150 \\
      \hline\hline
\end{tabular} 
\caption{Summary of the event yields  in the selected $WW$+1-jet events observed in data and estimated from signal and background contributions. The estimated event yields for the $WW$ signal are determined from simulated event samples which are scaled to a total cross section of $58.7^{+4.2}_{-3.8}$~pb. The estimated yields from diboson production  are  determined from simulated event samples whereas the yields of all other backgrounds are estimated using data-driven methods. The statistical and systematic uncertainties are shown separately. For reference, the numbers of observed, expected signal and background events for the $WW$+0-jet measurement~\cite{Collaboration:2010127} are also given.}\label{tab:Yields} 
\end{table}

\begin{figure}
\centering
\begin{subfigure}{.49\textwidth}
\includegraphics[width=\textwidth]{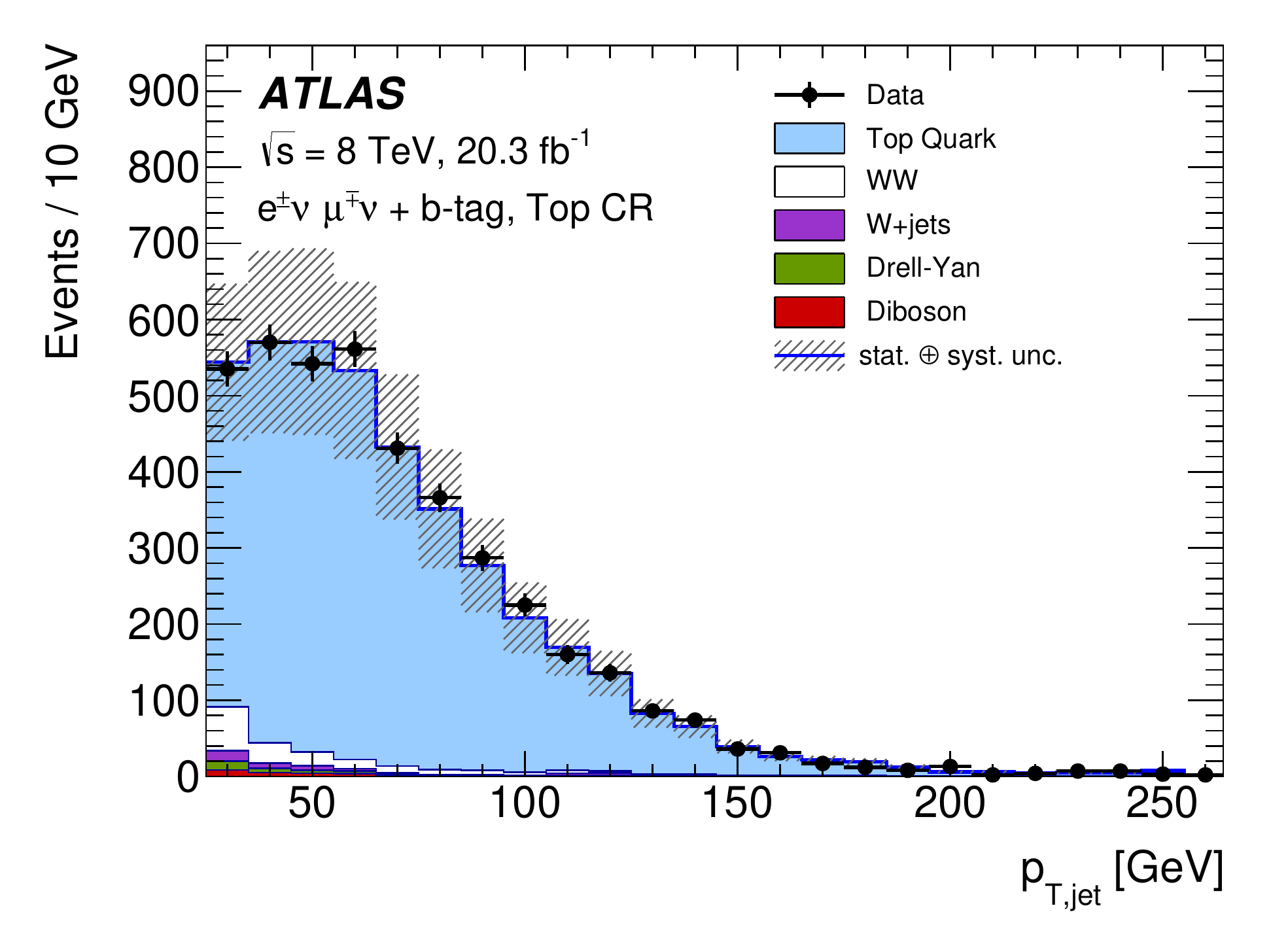}
\caption{}
\end{subfigure}
\begin{subfigure}{.49\textwidth}
\includegraphics[width=\textwidth]{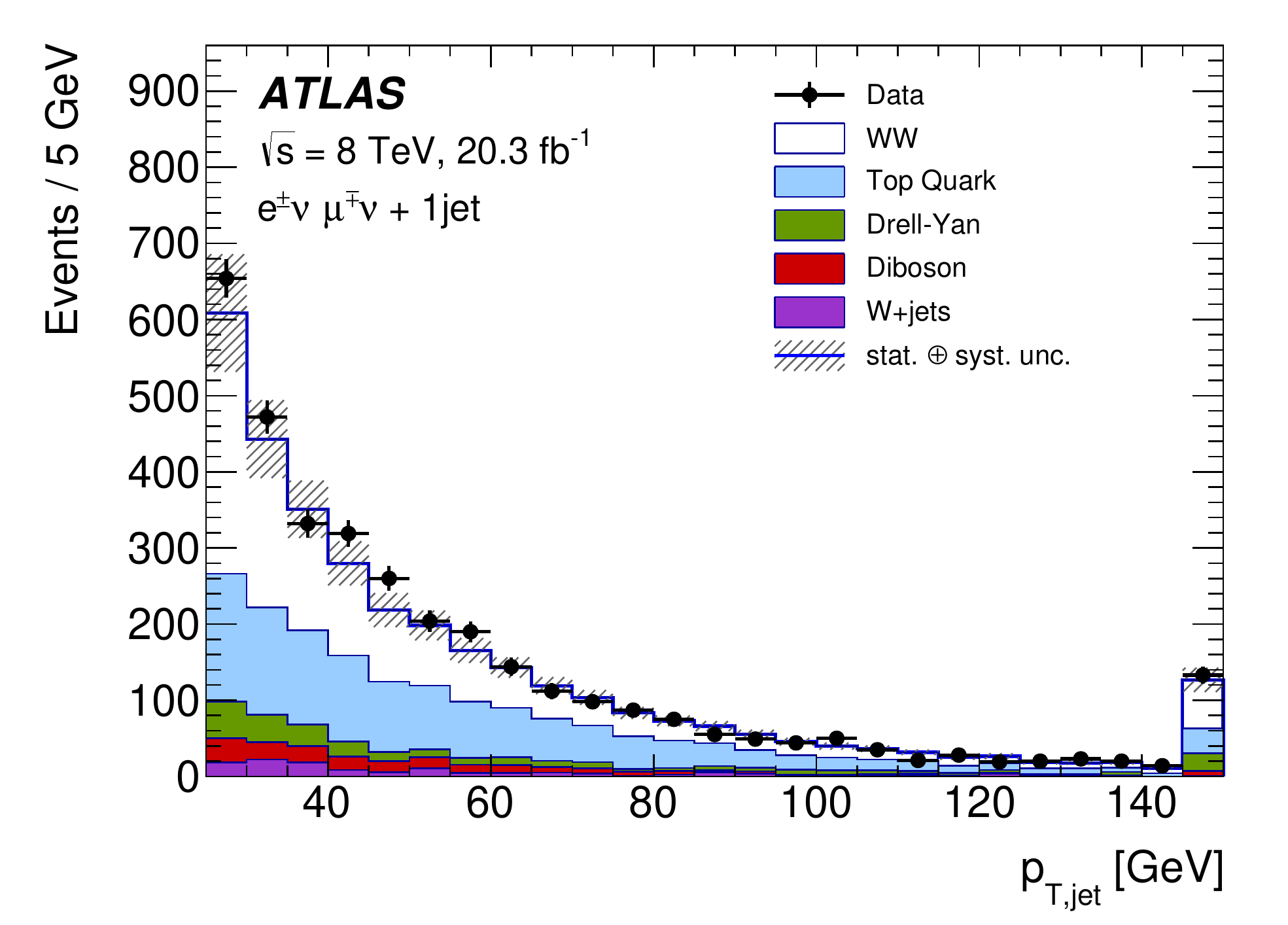}
\caption{}
\end{subfigure}
\caption{(a) Distributions of the transverse momentum of the selected jet in the control region enriched in events from top-quark production.  The sum in quadrature of statistical, experimental and theoretical uncertainties in the MC prediction are shown as a hatched band. (b) Distributions of the transverse momentum of the selected jet after final event selection. Data are shown together with the yields from $WW$ signal as estimated from simulated event samples  which are scaled to a total cross section of $58.7^{+4.2}_{-3.8}$~pb, and the estimated background contributions. The sum in quadrature of statistical, experimental and theoretical uncertainties is shown as a hatched band. In both figures the last bin of the distribution is an overflow bin.}
\label{fig:invmass-jetpta}
\label{fig:invmass-jetptb}
\label{fig:invmass-jetpt}
\end{figure}

\section{Cross-section measurement}
\label{sec:xsec}

The cross section for $WW$ production in the $e\mu$ final state with  exactly one jet is measured. 
The definition of the fiducial phase space is derived from the selection applied to reconstructed events.
Leptons are recombined with any final-state photons from QED radiation within a surrounding cone of size $\Delta R = 0.1$, to form so-called `dressed leptons'. Furthermore, electrons and muons are required to be oppositely charged and to originate directly from $W$ decays. 
The same selection requirements on transverse momentum and pseudorapidity as at reconstruction level are applied to the dressed leptons. 
Stable particles with a lifetime $\tau>30$~ps, excluding muons and neutrinos, are used to form particle-level jets using the anti-$k_t$ algorithm with a radius parameter of $R=0.4$. They are selected if $\pt>25$~\GeV\ and $\left|\eta\right| < 4.5$. To remove jets originating from electrons, jets which are a distance $\Delta R < 0.3$ from any electron from $W$ decays selected as detailed above are ignored. The four-momentum sum of the neutrinos originating from the $W$ boson decays is used for the calculation of both \ptmiss\ and \metrel\ at generator level. 

The number of selected $WW$ candidate events with exactly one  associated jet may receive contributions from events with different jet multiplicities due to the detector resolution. 
After subtracting the background contributions, $N_{\mathrm{b}}$, from the number of  observed events, $N_{\mathrm{obs}}$, the observed signal yield, $N_{\mathrm{s}}=N_{\mathrm{obs}}-N_{\mathrm{b}}$, is corrected for detector inefficiencies, resolution and jet migration effects using a correction matrix $R_{ij}$. The correction matrix also accounts for jets originating from pileup which increase the expected signal yield by 5\%.
It is evaluated using simulated $WW$ event samples as the ratio of the number of events reconstructed in jet-bin $i$ and generated in jet-bin $j$, $N_{\text{gen }j}^{\text{reco }i}$, to the number of events generated in the fiducial volume  with $j$ associated jets, $N_{\mathrm{gen }j}^{\mathrm{fid}}$:

\begin{equation}
R_{ij}=\frac{N_{\text{gen }j}^{\text{reco }i}}{N_{\mathrm{gen }j}^{\mathrm{fid}}}
\label{sec:xsec:defRij}
\end{equation}
where all jet multiplicities $j>1$ are contained in $N_{\text{gen }j}^{\text{reco }i}$ in the jet-bin corresponding to $j=1$ to account for migrations into the event sample.

Electrons and muons from non-prompt $\tau$-lepton decays are accounted for in the numerator of Eq.~(\ref{sec:xsec:defRij}) but not in the denominator, which effectively removes the contribution of $W\rightarrow\tau\nu$ decays. This allows a definition of  the fiducial region for prompt decays of $W$ bosons into electrons and muons only. While the calculation of the total $pp\rightarrow W^+W^-$ cross section at NNLO does not include $b$-quarks, such events can occur in the simulated event samples from gluon splitting, $g\rightarrow b\bar{b}$. The veto on identified $b$-jets affects these contributions in the calculation of the correction matrix $R_{ij}$. The effect on the measured cross section is less than 1\%. 
The values of the matrix $R_{ij}$ are given in Table~\ref{tab:xsec:cwwfactorssplit_paperyeah} together with their total uncertainties. Events reconstructed with the wrong jet multiplicity cause non-zero values for $R_{ij}$ with $i\neq j$.

\begin{table}
\centering
    \sisetup{round-mode = figures, 
      table-number-alignment = center }
\begin{tabular}{ l|
  S[round-precision = 3,table-format = 1.3]
  S[round-precision = 3,table-format = 1.3] 
|
  S[round-precision = 3,table-format = 1.3]
  S[round-precision = 3,table-format = 1.3]
||  
  S[round-precision = 3,table-format = 1.3]
}
\hline
\hline
{\small $R_{ij}$  ($i=n_{\mathrm{jets}}^{\mathrm{reco}}$, $j=n^{\mathrm{gen}}_{\mathrm{jets}}$)} & \multicolumn{1}{c}{$R_{00}$} & \multicolumn{1}{c  |}{$R_{01}$} & \multicolumn{1}{c}{$R_{10}$} & \multicolumn{1}{c||}{$R_{11}$} &\multicolumn{1}{c}{$A_{WW}$}\\
\hline

$q\bar{q}\rightarrow W^+W^-$              & 0.501429 & \multicolumn{1}{c|}{\num[round-precision=2]{0.0355441}} & \multicolumn{1}{c}{\num[round-precision=2]{0.0498351}} & 0.457693&  0.327\\
$gg\rightarrow W^+W^-$              & 0.502409 & \multicolumn{1}{c|}{\num[round-precision=2]{0.0614278}} & \multicolumn{1}{c}{\num[round-precision=2]{0.0673433}} & 0.449625&   0.447\\
$gg\rightarrow H\rightarrow W^+W^-$ & 0.410045 & \multicolumn{1}{c|}{\num[round-precision=2]{0.0348924}}& \multicolumn{1}{c}{\num[round-precision=2]{0.0553272}} & 0.423407&   0.169\\
\hline                                                                           
Total $WW$                      & 0.498504 & \multicolumn{1}{c|}{\num[round-precision=2]{0.036503}} & \multicolumn{1}{c}{\num[round-precision=2]{0.0505977}} & 0.455691&0.319\\  
Uncertainty                & \multicolumn{1}{l}{\num[round-precision=1]{4.3}\%}  & \multicolumn{1}{l|}{\num[round-precision=2]{45.2}\%}& \multicolumn{1}{l}{\num[round-precision=2]{24.3}\%}& \multicolumn{1}{l||}{\num[round-precision=1]{6.3}\%}&   \multicolumn{1}{l}{\num[round-precision=2]{4.9}\%} \\ 

\hline
\hline
\end{tabular}
\caption{Numerical values of the correction matrix $R_{ij}$ which accounts for the full detector efficiency migrations between jet bins, and the  factor $A_{WW}$ which accounts for the extrapolation from the $WW$+$\leq$1-jet final state to the total phase space. For both variables the total uncertainties are shown.}
\label{tab:xsec:cwwfactorssplit_paperyeah}
\end{table}

The fiducial $WW$ cross section in jet-bin $j$ is given by the measured signal yields in jet-bins $i=0,1$:

\begin{equation}
\sigma^{\mathrm{fid},j}_{WW}= \frac{1}{\mathcal{L}}\sum_{i=0}^{1} R_{ij}^{-1}N^i_{\mathrm{s}},
\label{eq:xsec:xsecdef:paper}
\end{equation}
where $\mathcal{L}$ is the integrated luminosity and $N^i_{\mathrm{s}}$ the background-subtracted events yield in jet bin $i$. 
The cross sections for $WW$ production  with zero and one associated jet are extracted simultaneously using a profile likelihood fit~\cite{Cranmer:2012sba,Verkerke:2003ir} to data observed in 0-jet and 1-jet final states. Information from both the 0-jet final states from Ref.~\cite{Collaboration:2010127} and 1-jet final states are used, where systematic uncertainties are added to the likelihood function as nuisance parameters and treated as correlated between 0-jet and 1-jet final states.

The sum of the fiducial 0-jet and 1-jet cross sections is 
extrapolated to the total phase space by correcting for the acceptance $A_{WW}$ and the branching fraction $\mathcal{B}$ of $W\rightarrow\ell\nu$ decays:

\begin{equation}
\sigma^{\mathrm{tot}}_{WW} =\frac{\sigma^{\mathrm{fid},0}_{WW}+\sigma^{\mathrm{fid},1}_{WW}}{A_{WW}\cdot\mathcal{B}^2}.\label{eq:sigmatot}
\end{equation}
Here, the acceptance $A_{WW}$ is defined as the ratio of events generated in the $\leq1$-jet fiducial volume to all generated events. The acceptance correction factor is $A_{WW}=0.319$, which is roughly 40\% larger than for pure $WW$+0-jet final states~\cite{Collaboration:2010127}. The $W\rightarrow \ell\nu$, $\ell=e,\mu\text{ or }\tau$, branching fraction is $\mathcal{B}=0.1083$~\cite{alcaraz:2006mx}. 
 
\section{Systematic uncertainties}
\label{sec:systematics}

Systematic uncertainties arising from the limited knowledge of the event reconstruction efficiency and the determination  of the particle four-momenta are propagated to the measurement by varying the corresponding parameters in the calculation of the correction matrix $R_{ij}$. Uncertainties in the efficiency of the trigger and the selection of the leptons result in an uncertainty of  $\pm1.8\%$ in the fiducial cross section~\cite{Aad:2014rra,Aad:2014sca,ATLAS-CONF-2012-048,atlas-conf-2014-032,PERF-2013-05}. An uncertainty   of $\pm2.9\%$~\cite{atlas-conf-2014-046} is  attributed to the identification and rejection of jets containing $b$-hadrons.

Uncertainties in the jet energy scale and the jet energy resolution affect the matrix elements $R_{ij}$ especially for events with jets near the transverse momentum threshold of $\pt=25$~\GeV, resulting in uncertainties that can be as large as $\pm 40\%$ for $R_{ij}$ with $i\neq j$.
The effect on the $WW+$1-jet cross section is found to be $\pm4.2\%$  and $\pm 1.0\%$ from the jet energy scale and resolution~\cite{JES,ATLAS-CONF-2015-037}, respectively. 
The uncertainty due to $E_{\mathrm{T}}^{\mathrm{miss}}$ scale and resolution as well as $p_{\mathrm{T}}^{\mathrm{miss}}$ scale and resolution account for $\pm0.4\%$ in total~\cite{ATLAS-CONF-2014-019}. The uncertainty from the modelling of additional $pp$ interactions occurring in the same or nearby bunch crossings is less than $\pm0.6\%$. 

Uncertainties in the fiducial cross section due to the theoretical modelling of the correction matrix $R_{ij}$ are evaluated using alternative simulated $q\bar{q}\rightarrow W^+W^-$ event samples. The uncertainty due to the choice of generator and parton shower model is estimated by comparing simulated event samples generated with \textsc{Powheg+Pythia~8} and  with \textsc{MC@NLO+Jimmy}. The resulting uncertainty in the measured cross section is $\pm 2.4\%$. The effect of higher-order corrections is estimated by varying the renormalisation and factorisation scales simultaneously by factors of 0.5 and 2 and comparing the resulting correction matrices. The associated uncertainty in the measured 1-jet cross section amounts to $\pm0.5\%$. The uncertainty due to the choice of PDF is calculated according to Ref.~\cite{Botje:2011sn} and amounts to less than $\pm 0.1\%$. Accounting for migrations from higher jet multiplicities introduces uncertainties of $\pm2.1\%$. The uncertainty in the correction matrix due the relative normalisations of the different signal samples, \qqWW, non-resonant $gg$ and resonant $gg\rightarrow H$ production, is found to be negligible in comparison to other uncertainties. 

The extrapolation from the fiducial to the total phase space introduces additional uncertainties. These are assessed separately for the $q\bar{q}\rightarrow W^+W^-$,  non-resonant $gg\rightarrow W^+W^-$ and resonant $gg\rightarrow H\rightarrow W^+W^-$ processes and amount to $\pm 1.9\%$ for the MC generator and parton shower uncertainty evaluated as described above. The PDF-induced uncertainty is estimated to be $\pm 0.8\%$. The uncertainties due to potential contributions from higher-order effects are determined to be $\pm 4.0\%$ originating from the restriction to specific jet multiplicities. They are computed in the total phase space by considering the scale dependence of successive inclusive jet-binned cross sections to be uncorrelated~\cite{Stewart:2011cf}. The scale dependence of the remaining selection criteria is assessed without applying any jet requirements and is found to be $\pm 0.2\%$.

\section{Results}
\label{sec:results}
The cross section for $WW$+1-jet production in the fiducial region is measured to be:

\begin{equation}
\sigma^{\text{fid,1-jet}}_{WW}=\xsecfidIjval\pm\xsecfidIjstat\left(\text{stat}\right)\pm\xsecfidIjsystdn\left(\text{syst}\right)\pm\xsecfidIjlumidn\left(\text{lumi}\right)~\text{fb}.
\end{equation}

The total relative uncertainty of the measured value is $\pm15\%$ and correlated with the uncertainty of the fiducial $WW$+0-jet cross section of $\sigma^{\text{fid,0-jet}}_{WW}=374 \pm7\left(\text{stat}\right)^{+25}_{-23}\left(\text{syst}\right)^{+8}_{-7}\left(\text{lumi}\right)$~fb presented in Ref.~\cite{Collaboration:2010127}. The correlation coefficient between the total uncertainties of the 0- and the 1-jet fiducial measurements is found to be $\rho=\rhoOIjet$. The  measured cross sections and uncertainties can be used to compute a cross section defined in the fiducial  $WW$+$\leq1$-jet region:

\begin{equation}\sigma^{\mathrm{fid,}\leq\mathrm{1-jet}}_{WW}=\xsecfidOIjval\pm\xsecfidOIjstat\left(\mathrm{stat}\right)\pm\xsecfidOIjsyst\left(\mathrm{syst}\right)\pm\xsecfidOIjlumi\left(\mathrm{lumi}\right)\mathrm{fb}.\end{equation}
Uncertainties causing migrations of events between jet bins are significantly reduced when comparing the fiducial $WW$+0-jet cross section and the $WW$+$\leq$1-jet cross section. The previously dominant experimental uncertainty in the jet energy scale is reduced by a factor of 2.5  by extending the measurement to include 1-jet final states.

Additional uncertainties introduced by the rejection of $b$-jets and increased uncertainties in the estimation of  background contributions cause the overall experimental uncertainty to be lower by only 18\%. 

The ratio of jet-binned fiducial cross sections $R_1$ is measured to be:

\begin{equation}
R_1=\sigma^{\text{fid,1-jet}}_{WW}\left/\sigma^{\text{fid,0-jet}}_{WW}\right.=0.36\pm0.05
\end{equation}
and allows a test of theoretical calculations without knowing  the total cross section.

Theoretical predictions of the fiducial cross sections are obtained by combining three separate theoretical calculations of the total cross sections with their respective acceptance correction factors $A_{WW}$. These factors are calculated using the simulated event samples generated at lower order in the perturbative expansion for the three separate processes contributing to $WW$ production.

The theoretical calculation of $pp\rightarrow W^+W^-$ to order $\mathcal{O}(\alphas^2)$~\cite{Gehrmann:2014fva} is used, which formally includes the loop-induced $gg$ contribution at order $\mathcal{O}(\alphas^2)$. This $gg$ contribution is subtracted and replaced by a calculation of the $gg$ loop-process to order $\mathcal{O}(\alphas^3)$~\cite{Caola:2015rqy} instead. To this non-resonant $WW$ prediction, the prediction for resonant $WW^*$ production via a Higgs boson with a subsequent decay into two $W$ bosons at order $\mathcal{O}(\alphas^4)$~\cite{HiggsReport4} is added to yield the total cross-section prediction of  $65.0^{+1.2}_{-1.1}$~pb,\footnote{The prediction for the total cross section is slightly larger than the one cited in Ref.~\cite{Collaboration:2010127} due to the inclusion of the higher-order calculation of the loop-induced $gg$ processes and the use of an alternative scale choice in the calculation of the $q\bar{q}\rightarrow W^+W^-$ process.} where the contributions from resonant and non-resonant $gg\rightarrow W^+W^-$ production amount to 6.4\% and 4.2\% of the total cross section, respectively. 
Theoretical uncertainties in the acceptance are assigned as described in Section~\ref{sec:systematics}. The approximate theoretical fiducial cross sections are found to be:
\begin{eqnarray}
  \sigma^{\text{fid,1-jet}}_{WW}             &=&141\pm30~\text{fb}\\
  \sigma^{\text{fid,}\leq\mathrm{1-jet}}_{WW}&=&487\pm22~\text{fb}.
\end{eqnarray}
A comparison of the measured and predicted fiducial cross sections is given in Figure~\ref{fig:finalresult:fidxsecs}(a). While the fiducial $WW$+0-jet cross section was measured slightly higher than the theoretical prediction, the fiducial $WW$+1-jet and $WW$+$\leq$1-jet cross-section measurements agree well with the theoretical prediction. 

The ratio of the jet-binned fiducial cross sections $R_1$ measured in data is compared to
several theoretical predictions in Figure~\ref{fig:finalresult:yeah:paper}(b). All theoretical values agree well with the measurement within uncertainties.  The first two theoretical predictions are taken from either the \textsc{Powheg+Pythia~8} or the \textsc{MC@NLO+Jimmy} \qqWW\ samples. The
theoretical uncertainty in these predictions is assessed by varying the
renormalisation and factorisation scales independently by factors of
0.5 and 2 with the constraint $0.5 < \mu_{\mathrm F}/\mu_{\mathrm R} < 2$. The contributions from
resonant and non-resonant $gg\rightarrow W^+W^-$  production are taken in both cases from the respective \textsc{Powheg+Pythia~8} and \ggtwoww\ samples, which increase the prediction for $R_1$ due to more initial-state radiation from gluons than quarks. The full effect of omitting the 
$gg\rightarrow W^+W^-$  contributions is assigned as further theoretical uncertainty.
To investigate resummation effects, a third prediction is obtained 
from the \qqWW\ and $gg\rightarrow W^+W^-$ samples as discussed above, but with the
\textsc{Powheg+Pythia~8} \qqWW\ sample reweighted to reproduce the  $p_{\mathrm{T},WW}$  distribution as predicted by the NLO+NNLL calculation in Ref.~\cite{Meade:2014fca}. In addition to renormalisation and factorisation scales,
 the resummation scale is  varied here.
Finally, predictions for $R_1$ are obtained by using recent fixed-order
calculations for the \qqWW\ and non-resonant $gg\rightarrow W^+W^-$ processes from
\textsc{Matrix} at NNLO~\cite{Grazzini:2016ctr} and \textsc{MCFM} at NLO, where the latter uses the
implementations of inclusive $WW$ production~\cite{Campbell:2011bn} and $WW$+1-jet production~\cite{Campbell:2015hya}. These programs allow the application of the fiducial lepton and
missing transverse momentum selections avoiding the use of acceptance
factors derived from lower-order  programs. Jets are clustered from
the final state partons using the anti-$k_t$ algorithm with the radius
parameter $R=0.4$. A correction for non-perturbative effects from
hadronisation and the underlying event is derived by comparing samples of \textsc{Madgraph}~\cite{Alwall:2014hca} using the CT10 PDF
interfaced with \textsc{Pythia~8} and the AU2 tune  with these effects enabled or
disabled. A systematic uncertainty is derived by interfacing the \textsc{Madgraph} samples
 with \textsc{Herwig++}~\cite{Bahr:2008pv} and the AUET2 tune.
The
renormalisation and factorisation scales for the \textsc{Matrix} and \textsc{MCFM}
predictions are set to $\mu_{\mathrm R}=\mu_{\mathrm F}=m_W$ and an uncertainty is obtained by varying
those independently by factors of 0.5 and 2 with the constraint $0.5 <
\mu_{\mathrm F}/\mu_{\mathrm R} < 2$. In both of these calculations, the non-resonant $gg\rightarrow W^+W^-$
production only contributes in the denominator of $R_1$. Contributions
from resonant $gg\rightarrow H\rightarrow W^+W^-$ production are included using event samples simulated with \textsc{Powheg+Pythia~8}.

\begin{figure}
\centering
\begin{subfigure}{0.49\textwidth}
\includegraphics[width=\textwidth]{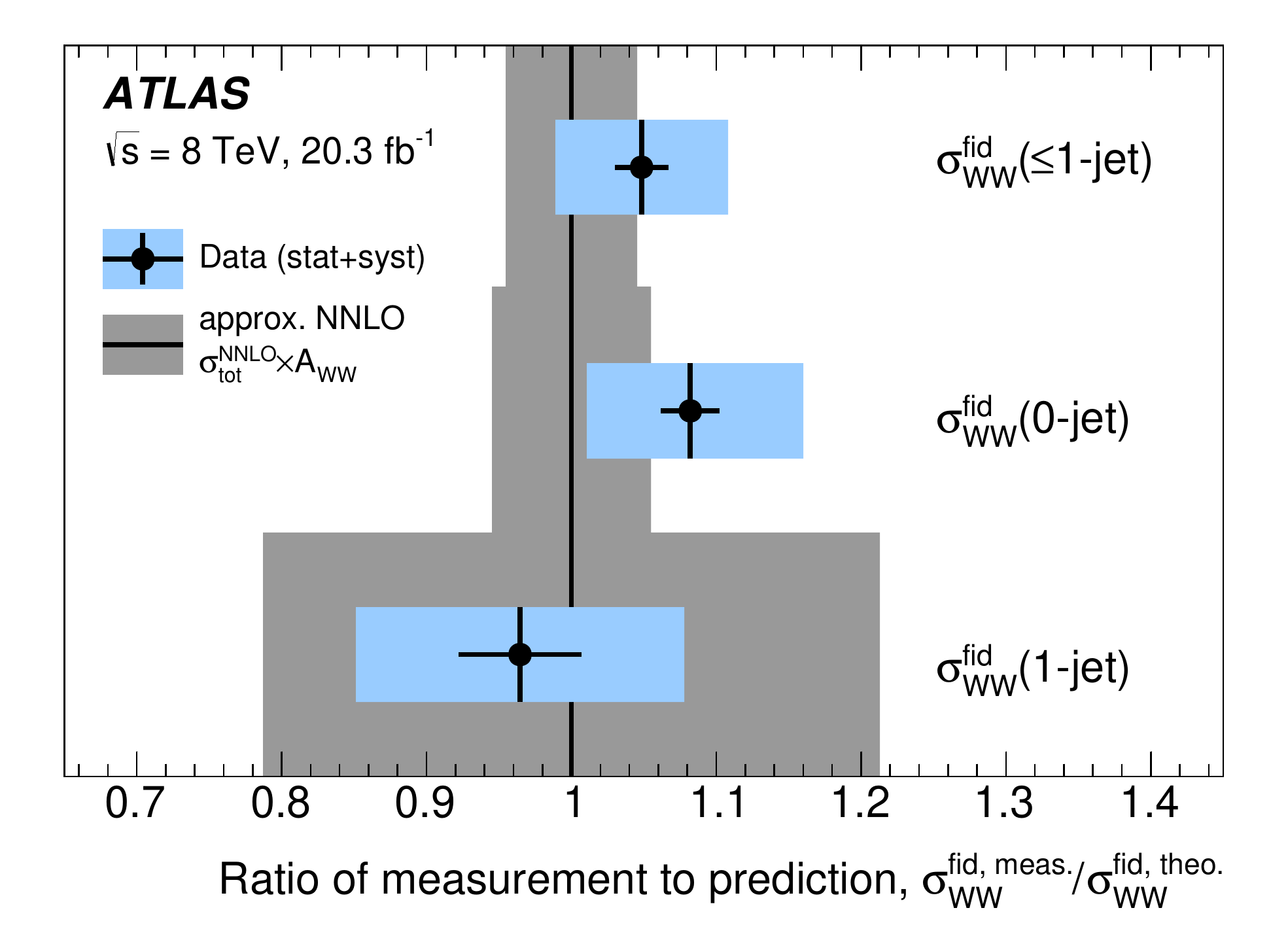}
\caption{}
\end{subfigure}
\begin{subfigure}{0.49\textwidth}
\includegraphics[width=\textwidth]{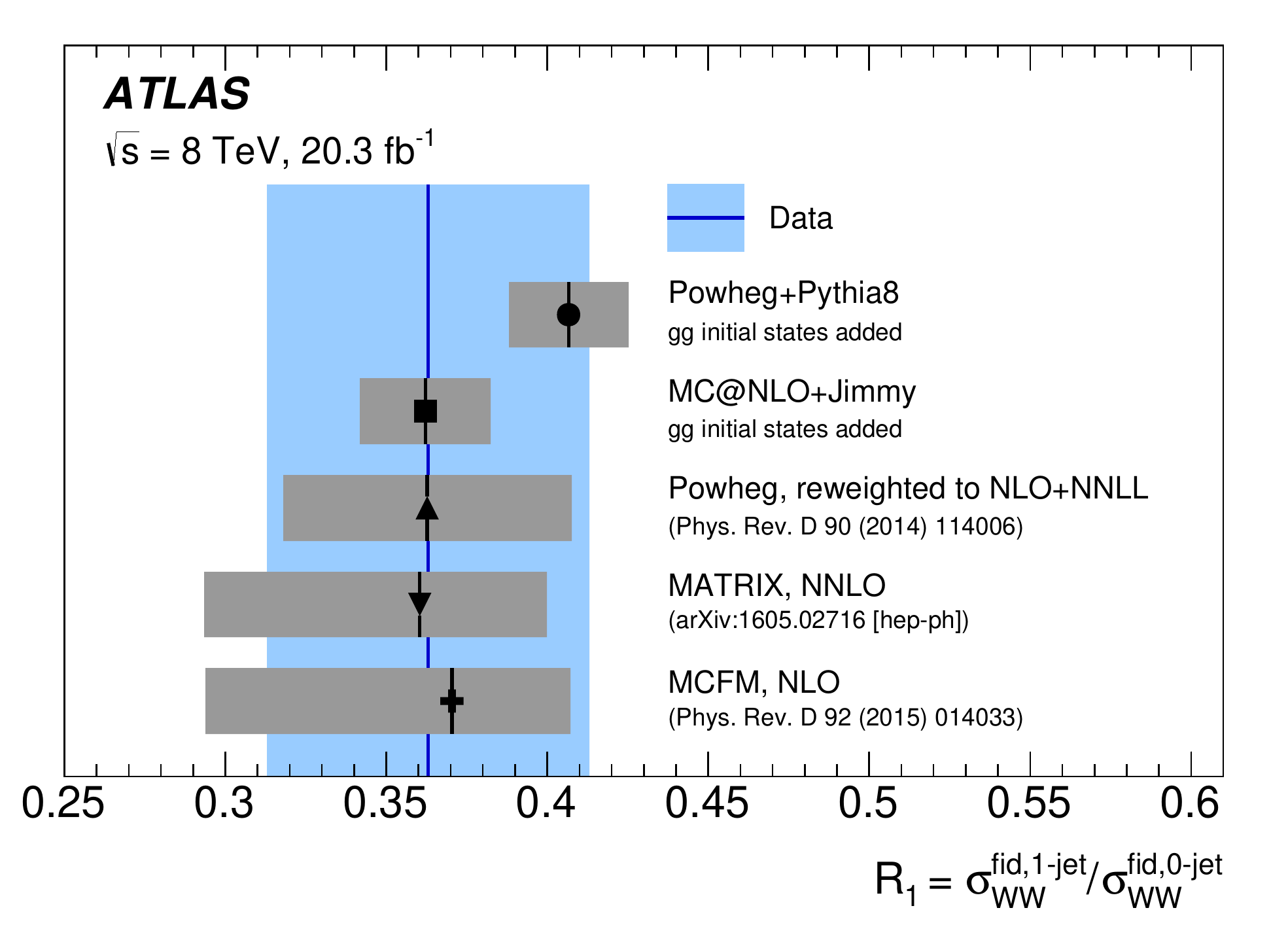}
\caption{}
\end{subfigure}
\caption{(a) Comparison of the measured cross sections in the 0-jet, 1-jet and $\leq$1-jet fiducial regions. The ratio of the measured cross sections to their respective theoretical prediction is shown. The theoretical predictions were obtained by multiplying the total cross section of $65.0^{+1.2}_{-1.1}$~pb with the total acceptance obtained by combining the acceptance correction factors $A_{WW}$ for the $WW$ processes according to their contribution. (b) Jet-binned fiducial cross-section ratio $R_1$ measured in data and compared to theoretical predictions. The values are obtained for two different \qqWW\  generators and by reweighting \textsc{Powheg+Pythia~8} to a resummation calculation at NLO+NNLL. Contributions from resonant and non-resonant $gg\rightarrow W^+W^-$ production are added to all three theoretical values. Fixed-order calculations at NNLO using \textsc{Matrix}~\cite{Grazzini:2016ctr} and at NLO using \textsc{MCFM}~\cite{Campbell:2011bn,Campbell:2015hya} are also shown, where contributions from $gg\rightarrow H\rightarrow W^+W^-$ production are added using simulated \textsc{Powheg+Pythia~8} samples.
For the measured cross sections in (a) and (b) the combined statistical and systematic uncertainties are shown as a blue band. When statistical uncertainties are given they are indicated as horizontal error bars. The uncertainties in theoretical cross sections are shown as a grey band.
}
\label{fig:finalresult:fidxsecs}
\label{fig:finalresult:yeah:paper}
\end{figure}

The total $WW$ cross section is extrapolated from the fiducial $WW$+$\leq$1-jet cross section using Eq.~(\ref{eq:sigmatot}) and found to be:

\begin{equation}
\sigma^{\mathrm{tot}}_{WW}  = \xsectotval\pm\xsectotstat(\text{stat})\pm\xsectotsyst(\text{syst})\pm\xsectottheo(\text{theo})\pm\xsectotlumi(\text{lumi})\text{ pb.}
\end{equation}

The result presented here is 12\% more precise than the previous ATLAS measurement based on $WW$+0-jet candidate events only~\cite{Collaboration:2010127} due to smaller experimental uncertainties in the fiducial $WW$+$\leq$1-jet cross-section measurement. 
The measured cross section is compatible with the theoretical prediction of $65.0^{+1.2}_{-1.1}$~pb. 

\section{Conclusion}
\label{sec:conclusion}
\label{sec:summary} 
The production of $W$ boson pairs in association with a hadronic jet was studied in $pp$ collisions at a centre-of-mass energy of $\sqrt{s}=8$~\TeV\ using data with an integrated luminosity of 20.3~\ifb\ collected by the ATLAS detector at the LHC. The analysis extends a previous analysis to final states with one jet. The fiducial $WW$+1-jet cross section is measured to be $\xsecfidIjval\pm\xsecfidIjtoterrdn$~fb within the fiducial volume defined by the kinematic requirements placed in the analysis. It is found to be in very good agreement with the theoretical prediction obtained by combining the total cross-section calculations of $q\bar{q}\rightarrow W^+W^-$ at $\mathcal{O}\left(\alphas^2\right)$, non-resonant $gg\rightarrow W^+W^-$ at $\mathcal{O}\left(\alphas^3\right)$, and resonant $gg\rightarrow W^+W^-$ at $\mathcal{O}\left(\alphas^4\right)$ and multiplying them with their respective acceptance factor $A_{WW}$. Similarly, the measured fiducial $WW$+$\leq$1-jet cross section of $\xsecfidOIjval\pm\xsecfidOIjtoterr$~fb agrees within the uncertainty with the prediction. The fiducial $WW$+$\leq$1-jet cross section is extrapolated to the total phase space, yielding a measurement of the total $pp\rightarrow W^+W^-$ cross section of $\xsectotval\pm\xsectottot$~pb. This result is compared to the  highest-order theory calculation available of $65.0\pm1.2$~pb. 

The total cross section extrapolated from the $\leq$1-jet fiducial volume is in better agreement with the theory calculation than the total cross section extrapolated from the 0-jet fiducial volume. The uncertainty is improved by 12\%.

To investigate further how well current predictions are able to describe the relative contributions of these exclusive jet cross sections, the ratio of the fiducial $WW$+1-jet to the fiducial $WW$+0-jet cross section, $R_1$, is determined to be $0.36\pm0.05$ and compared to various theoretical predictions, which are all found to agree  with the measurement within the uncertainties.


\section*{Acknowledgements}


We thank CERN for the very successful operation of the LHC, as well as the
support staff from our institutions without whom ATLAS could not be
operated efficiently.

We acknowledge the support of ANPCyT, Argentina; YerPhI, Armenia; ARC, Australia; BMWFW and FWF, Austria; ANAS, Azerbaijan; SSTC, Belarus; CNPq and FAPESP, Brazil; NSERC, NRC and CFI, Canada; CERN; CONICYT, Chile; CAS, MOST and NSFC, China; COLCIENCIAS, Colombia; MSMT CR, MPO CR and VSC CR, Czech Republic; DNRF and DNSRC, Denmark; IN2P3-CNRS, CEA-DSM/IRFU, France; GNSF, Georgia; BMBF, HGF, and MPG, Germany; GSRT, Greece; RGC, Hong Kong SAR, China; ISF, I-CORE and Benoziyo Center, Israel; INFN, Italy; MEXT and JSPS, Japan; CNRST, Morocco; FOM and NWO, Netherlands; RCN, Norway; MNiSW and NCN, Poland; FCT, Portugal; MNE/IFA, Romania; MES of Russia and NRC KI, Russian Federation; JINR; MESTD, Serbia; MSSR, Slovakia; ARRS and MIZ\v{S}, Slovenia; DST/NRF, South Africa; MINECO, Spain; SRC and Wallenberg Foundation, Sweden; SERI, SNSF and Cantons of Bern and Geneva, Switzerland; MOST, Taiwan; TAEK, Turkey; STFC, United Kingdom; DOE and NSF, United States of America. In addition, individual groups and members have received support from BCKDF, the Canada Council, CANARIE, CRC, Compute Canada, FQRNT, and the Ontario Innovation Trust, Canada; EPLANET, ERC, FP7, Horizon 2020 and Marie Sk{\l}odowska-Curie Actions, European Union; Investissements d'Avenir Labex and Idex, ANR, R{\'e}gion Auvergne and Fondation Partager le Savoir, France; DFG and AvH Foundation, Germany; Herakleitos, Thales and Aristeia programmes co-financed by EU-ESF and the Greek NSRF; BSF, GIF and Minerva, Israel; BRF, Norway; Generalitat de Catalunya, Generalitat Valenciana, Spain; the Royal Society and Leverhulme Trust, United Kingdom.

The crucial computing support from all WLCG partners is acknowledged gratefully, in particular from CERN, the ATLAS Tier-1 facilities at TRIUMF (Canada), NDGF (Denmark, Norway, Sweden), CC-IN2P3 (France), KIT/GridKA (Germany), INFN-CNAF (Italy), NL-T1 (Netherlands), PIC (Spain), ASGC (Taiwan), RAL (UK) and BNL (USA), the Tier-2 facilities worldwide and large non-WLCG resource providers. Major contributors of computing resources are listed in Ref.~\cite{ATL-GEN-PUB-2016-002}.



\printbibliography

\clearpage

\newpage 
\begin{flushleft}
{\Large The ATLAS Collaboration}

\bigskip

M.~Aaboud$^\textrm{\scriptsize 136d}$,
G.~Aad$^\textrm{\scriptsize 87}$,
B.~Abbott$^\textrm{\scriptsize 114}$,
J.~Abdallah$^\textrm{\scriptsize 8}$,
O.~Abdinov$^\textrm{\scriptsize 12}$,
B.~Abeloos$^\textrm{\scriptsize 118}$,
R.~Aben$^\textrm{\scriptsize 108}$,
O.S.~AbouZeid$^\textrm{\scriptsize 138}$,
N.L.~Abraham$^\textrm{\scriptsize 152}$,
H.~Abramowicz$^\textrm{\scriptsize 156}$
% ...
\bigskip
\\
% ...
$^{8}$ Department of Physics, The University of Texas at Arlington, Arlington TX, United States of America\\
% ...
$^{12}$ Institute of Physics, Azerbaijan Academy of Sciences, Baku, Azerbaijan\\
% ...
$^{87}$ CPPM, Aix-Marseille Universit{\'e} and CNRS/IN2P3, Marseille, France\\
% ...
$^{108}$ Nikhef National Institute for Subatomic Physics and University of Amsterdam, Amsterdam, Netherlands\\
% ...
$^{114}$ Homer L. Dodge Department of Physics and Astronomy, University of Oklahoma, Norman OK, United States of America\\
% ...
$^{118}$ LAL, Univ. Paris-Sud, CNRS/IN2P3, Universit{\'e} Paris-Saclay, Orsay, France\\
% ...
$^{136}$ $^{(a)}$ Facult{\'e} des Sciences Ain Chock, R{\'e}seau Universitaire de Physique des Hautes Energies - Universit{\'e} Hassan II, Casablanca; $^{(b)}$ Centre National de l'Energie des Sciences Techniques Nucleaires, Rabat; $^{(c)}$ Facult{\'e} des Sciences Semlalia, Universit{\'e} Cadi Ayyad, LPHEA-Marrakech; $^{(d)}$ Facult{\'e} des Sciences, Universit{\'e} Mohamed Premier and LPTPM, Oujda; $^{(e)}$ Facult{\'e} des sciences, Universit{\'e} Mohammed V, Rabat, Morocco\\
% ...
$^{138}$ Santa Cruz Institute for Particle Physics, University of California Santa Cruz, Santa Cruz CA, United States of America\\
% ...
$^{152}$ Department of Physics and Astronomy, University of Sussex, Brighton, United Kingdom\\
% ...
$^{156}$ Raymond and Beverly Sackler School of Physics and Astronomy, Tel Aviv University, Tel Aviv, Israel\\
% ...
$^{a}$ Also at Department of Physics, King's College London, London, United Kingdom\\
$^{b}$ Also at Institute of Physics, Azerbaijan Academy of Sciences, Baku, Azerbaijan\\
$^{c}$ Also at Novosibirsk State University, Novosibirsk, Russia\\
$^{d}$ Also at TRIUMF, Vancouver BC, Canada\\
$^{e}$ Also at Department of Physics {\&} Astronomy, University of Louisville, Louisville, KY, United States of America\\
% ...
\end{flushleft}



\end{document}